# Low-loss Materials for high Q-factor Bragg Reflector Resonators

Jean-Michel le Floch, Michael E.Tobar

[1] School of Physics, University of Western Australia, 35 Stirling Hwy, Crawley, WA, 6009

Dominique Cros

XLIM - UMR CNRS n°6172 - Faculté des Sciences et Techniques - 123, avenue A. Thomas, 87060 Limoges Cedex, France

Jerzy Krupka

[3] Institute of Microelectronics and Optoelectronics, University of Technology, Koszykowa 75, Warsaw, Poland

A Bragg resonator uses dielectric plates within a metallic cavity to confine the energy within a central free space region. The importance of the permittivity is shown with a better Q-factor possible using higher permittivity materials of larger intrinsic dielectric losses. This is because the electric energy in the reflectors decreases proportionally to the square root of permittivity and the coupling to the metallic losses decrease linearly. In a sapphire resonator with a single reflector pair a Q-factor of $2.34 \times 10^5$ is obtained, which may be improved on by up to a factor of 2 using higher permittivity materials.

Bragg reflector resonators obtain high Q-factors by confining most of the energy in a central low-loss region (usually vacuum) using outer-layered dielectric materials loaded in a metallic cavity. The dielectric and resistive losses are decreased due to the reduced electromagnetic energy in the dielectric layers and at the cavity surface. So far the highest Q-factors ($\sim 7 \times 10^5$) have been obtained in Bragg resonators with two pairs of sapphire reflectors due to its extremely low loss at microwave frequencies [1, 2]. In contrast traditional $TE_{01\delta}$ modes only obtain a Q-factor of order $4\text{-}5 \times 10^4$ [3,4]. In this work we investigate a range of alternative low-loss crystalline and non-crystalline dielectric materials [5] with the aim to obtain a higher Q-factor in a Bragg reflector resonator.

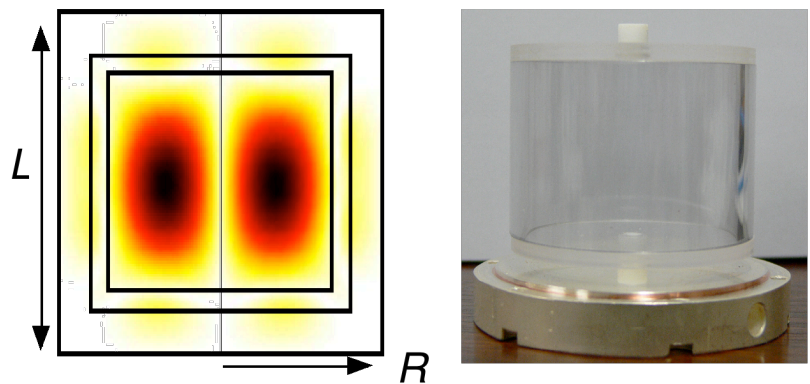

Figure 1. Right, photo of a single pair sapphire Bragg reflector resonator of frequency 9.7 GHz and Q-factor $2.34 \times 10^5$ at room temperature. Left, Electric field density plot ($E_\theta$) of the fundamental TE mode as calculated using the Method of Lines, showing the internal resonant region (free space), and outer anti-resonant layered region. For 10 GHz the size ranges between L= 60 – 100 mm, with R ~ 30 mm for AR ranging between 1.0 – 1.75.

To compare the efficiency of various materials a standard structure has been analyzed made from a single Bragg reflector pair in the radial and axial regions as shown in figure 1. It has been shown that the near optimum solution may be obtained when we use a simple model to calculate the dimensions of the resonator, with a Q-factor of order $2\text{-}3 \times 10^5$ [6]. The simple model is briefly presented and used in this work to calculate the dimensions then a numerical technique based on the Method of Lines (MoL) was used to calculate the resonator properties. The MoL is a semi-analytical method with a rigorous formulation initially developed by mathematicians for the solution of differential equations [7].

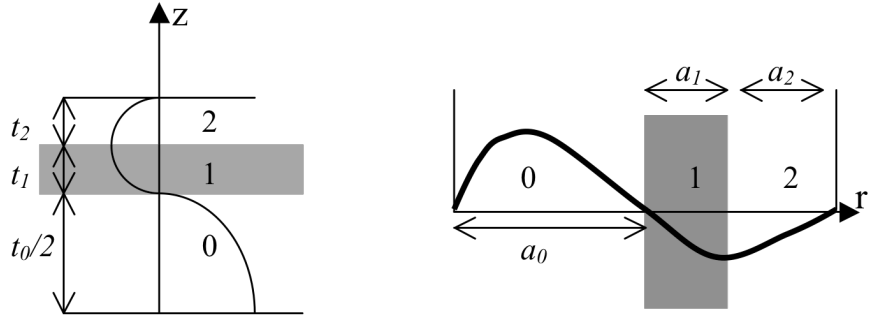

Figure 2: Schematic of the $E_\theta$ field patterns from the simple model, showing boundary conditions in the axial (left) and radial (right) directions. The gray regions 1 represent the dielectric reflector layer, while 2 is the free space reflector region. Region 0 is the central resonant region where the energy is confined, with the origin in the middle.

To design such structures, many methods of solving the electromagnetic boundary value problems have been developed. Previously, we have developed an approximate solution for a Bragg resonator, using a standard non-Maxwellian model procedure [6]. The method divides the resonator into sections then applies separation of variables technique for the $E_\theta$ component of field in each of the sections individually, while matching the boundary conditions between them (see figure 2). The structure is divided into the main resonance region of thickness $t_0$ and radius $a_0$ (region 0), and the Bragg reflector regions 1 and 2. The Bragg reflector region has 2-layers in the axial (thickness $t_1$ and $t_2$) and radial (thickness $a_1$ and $a_2$) directions. From solving the boundary conditions the following equations (1-3) are derived [6], and can be used to calculate the dimensions (the derivation is not repeated here) for a specific $AR$, $\varepsilon_r$ and resonant frequency, $f_0$, (related to the propagation constant $k_0$ by $k_0 = \frac{2\pi}{c} f_0$).

Solving the boundary conditions in the $r$-direction give (related to the roots of the first order Bessel function, $J_1(\chi)$, and its derivative):

$$a_0 = \frac{\gamma\, \chi_{1,1}}{k_0 \sqrt{\gamma^2 - 1}} \tag{1a}$$

$$a_1 = \frac{\gamma}{k_0 \sqrt{\varepsilon_r\, \gamma^2 - 1}} (\chi_{1,2} - \chi_{1,1}) \tag{1b}$$

$$a_2 = \frac{\gamma}{k_0 \sqrt{\gamma^2 - 1}} \left( \chi_{1,2} - \chi'_{1,2} \right) \tag{1c}$$

Solving the boundary conditions in the z-direction give:

$$t_0 = \frac{\gamma \pi}{k_0} \tag{2a}$$

$$t_1 = \frac{\pi}{2k_0 \sqrt{\varepsilon_r}} \tag{2b}$$

$$t_2 = \frac{\pi}{2k_0} \tag{2c}$$

Thus, the Aspect Ratio (*AR*) may be calculated to be:

$$AR = \frac{L/2}{R} = \frac{\pi \sqrt{\gamma^2 - 1}}{2} \left( \frac{\left(1 + \frac{1}{\gamma \sqrt{\varepsilon_r}} + \frac{1}{\gamma}\right)}{\chi_{1,1} + \chi_{1,2} - \chi'_{1,2} + \sqrt{\frac{\gamma^2 - 1}{\varepsilon_r \gamma^2 - 1}} \left( \chi'_{1,2} - \chi_{1,1} \right)} \right) \tag{3}$$

From the chosen *AR* and $\varepsilon_r$, one then calculates the coefficient $\gamma$ from (3), then from the selected $f_0$ all the dimensions from (1) and (2).

A dielectric material may be characterized by its complex permittivity $\varepsilon = \varepsilon_r - j\varepsilon''$. The loss tangent can be expressed from the complex permittivity as $\tan \delta = \frac{\varepsilon''}{\varepsilon_r}$. For the chosen resonant mode in a dielectric resonator, the Q-factor and frequency will depend on the above dielectric properties, as well as the dimensions, surface resistance (*Rs*) of the metallic enclosure, and specific mode properties. The mode properties are characterized by the Geometric-factor (*G*) and the electric energy filling factor in the dielectric (*Pe*).

The dimensions of the cavity (as calculated using the simple model) were entered in the MoL mesh for a range of $\varepsilon_r$ values between 10 and 200 and the *G*, *Pe* and $f_0$ calculated for each value of $\varepsilon_r$. The calculated $f_0$ using the MoL always gave the solution of 10 GHz to within 0.02 – 0.3 % as expected reconfirming the applicability of the simple model. The standard relationships between $Q_0$, $G$ and $Pe$ is given by:

$$Q_0^{-1} = Pe.tan\delta + \frac{Rs}{G} \quad (4)$$

Thus the required $tan\delta$ of the material, necessary to give an unloaded Q-factor ($Q_0$) of $3\times10^5$ (assuming the $Rs$ to be 26 m$\Omega$ the value of good silver plated copper [8] at 10GHz) allowed the numeric calculation of a locus in the permittivity versus loss tangent graph shown in figure 3 and can be calculated from (4) to be:

$$tan\delta\big|_{Q_0=3\times10^5} = \frac{3.333\times10^{-6} - 0.026/G_{BM}}{Pe_{BM}} \quad (5)$$

Here $Pe_{BM}$ and $G_{BM}$ are the electrical energy filling factor and G factor numerically calculated for the Bragg Mode (BM) at the specific value of $\varepsilon_r$.

**TABLE I**: Low-loss dielectric materials plotted in figure 2, with measured permittivities and loss tangents referred to 10 GHz [5]. The Q-factor of the structure in figure 1 is calculated using Method of Lines. For anisotropic materials we only give the perpendicular values, which is the only one relevant for the TE Bragg confined mode.

| No. | Material | Permittivity | $tan\delta$ [10 GHz] | Q-factor |
|---|---|---|---|---|
| 1 | CaTiO$_3$ [9] | 162 | $7.7\times10^{-4}$ | $4.7\times10^4$ |
| 2 | TiO$_2$ +0.05 mol% Fe [9] | 104 | $2.0\times10^{-4}$ | $1.3\times10^5$ |
| 3 | Pb$_{0.7}$Ca$_{0.3}$La$_{0.5}$(Mg$_{\frac{1}{2}}$Nb$_{\frac{1}{2}}$)O$_3$ [10] | 50 | $1.2\times10^{-4}$ | $1.5\times10^5$ |
| 4 | Ba[(Zn$_{0.6}$Co$_{0.4}$)$_{1/3}$Nb$_{2/3}$]O$_3$ [11] | 35.6 | $2.8\times10^{-5}$ | $3.9\times10^5$ |
| 5 | Ba(Mg$_{\frac{1}{2}}$Ta$_{2/3}$)O$_3$: 0.5mol% Ba(Mg$_{\frac{1}{2}}$W$_{\frac{1}{2}}$)O$_3$ [12] | 24.2 | $2.5\times10^{-5}$ | $3.4\times10^5$ |
| 5 | Ba(Mg$_{1/3}$Ta$_{2/3}$)O$_3$: BaSnO$_3$, BaWO$_4$ [12-15] | 24 | $2.3\times10^{-5}$ | $3.5\times10^5$ |
| 6 | 0.92(Mg$_{0.95}$Co$_{0.05}$)TiO$_3$-0.08CaTiO$_3$ [16] | 22.1 | $1.2\times10^{-5}$ | $4.7\times10^5$ |
| 7 | Al$_2$O$_3$ (TiO$_2$ doped) [17] | 10.5 | $8.4\times10^{-6}$ | $2.9\times10^5$ |
| 8 | Al$_2$O$_3$ Alumina [18,19] | 10.05 | $1.5\times10^{-5}$ | $2.3\times10^5$ |
| 9 | TiO$_2$ Rutile [20] | 85.7 | $1.5\times10^{-4}$ | $1.5\times10^5$ |
| 10 | LaGaO$_3$ [21] | 26 | $1.7\times10^{-5}$ | $4.5\times10^5$ |
| 11 | LaAlO$_3$ [22, 23] | 24 | $2.0\times10^{-5}$ | $3.9\times10^5$ |
| 12 | SrLaAlO$_4$ [24] | 16.85 | $1.6\times10^{-5}$ | $3.3\times10^5$ |
| 13 | YAG [24] | 10.6 | $1.2\times10^{-5}$ | $2.6\times10^5$ |
| 14 | AL$_2$O$_3$ Sapphire [24-26] | 9.935 | $8.5\times10^{-6}$ | $2.8\times10^5$ |

A range of materials at 10 GHz from the database given in [5] was compared with the calculated $tan\delta|_{Q_0=3\times10^5}$ locus from (5) and is shown in figure 3 and tabulated in table 1. As expected, crystalline sapphire (No. 14) and rutile-doped alumina (No. 7) are the lowest loss tangent materials and lie close to the curve, (Q-factor calculated to be ~ $2.8\times10^5$ and measured to be $2.34\times10^5$ ).

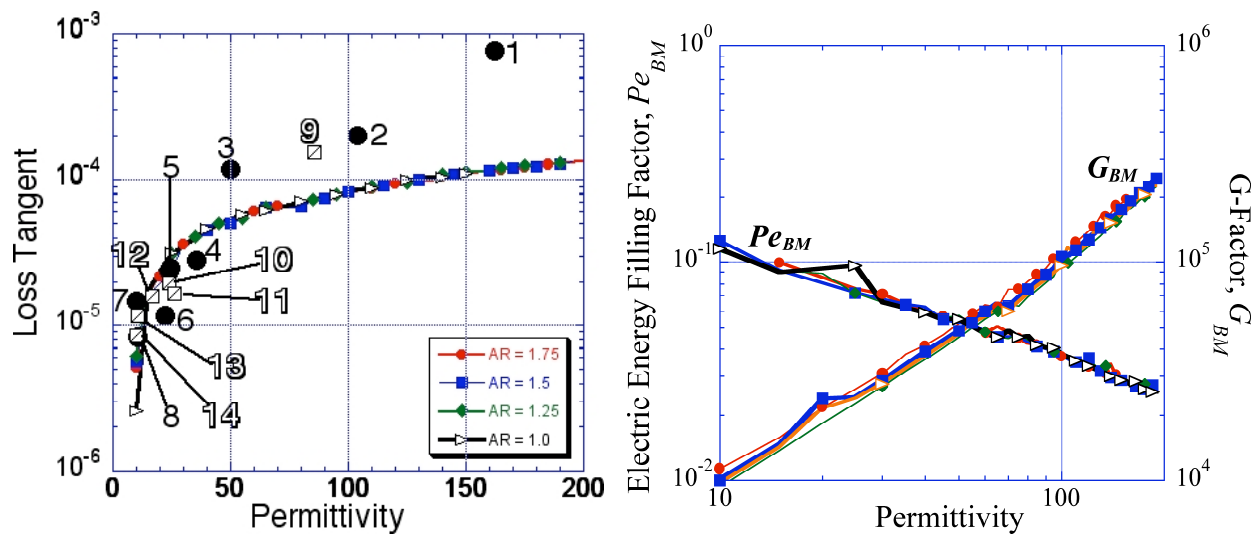

Figure 3. Left: Loss tangent versus permittivity showing the locus of $tan\delta|_{Q_0=3\times10^5}$ at 10 GHz as calculated using the MoL for the resonator design in figure 1 for a range of *AR* values. Fourteen low loss materials of varying permittivity are also plotted (see table I). The bold circles (1-8) represent non-crystalline material, while the white squares represent crystalline material. The materials below the locus have a Q-factor greater than $3\times10^5$, while the ones above have a lower Q-factor. Right: Electrical energy filling factor and mode G-factor of the as a function of permittivity as calculated using the MoL for a range of *AR* values. Results show an inverse square root and linear dependence respectively.

There is a range of crystalline and non-crystalline materials with higher loss and permittivity, which give higher Q-factors above $3\times10^5$, with only a small reduction in the thickness of the reflectors when compared to sapphire (thickness of the reflector is inversely proportional to the square root of the permittivity). This includes dielectric materials 4, 5 and 6 and the crystalline materials 10, 11 and 12 (see table 1). The machining of such structures should be no more difficult than sapphire or alumina [27], and in particular the dielectric materials can offer a cheaper alternative and a higher Q-factor.

However, the loss tangent of materials of very high permittivity (greater than 50 labeled 1-3 and 9), are not low enough to give a Q-factor greater than sapphire.

In conclusion, we have shown that an improved Q-factor may be obtained with a Bragg resonator by constructing the resonator from higher permittivity materials even if they have higher losses than sapphire. This occurs because of the smaller electric filling factor in the dielectric reflector and the higher *G*-factor of the mode for larger permittivity materials. This work was also extended to multi-layered Bragg reflector resonators, but is not presented here as the results give the same trends, except that Q-factors of better than $7 \times 10^5$ may be obtained.